\documentclass[graybox]{svmult}

\usepackage{mathptmx}       
\usepackage{helvet}         
\usepackage{courier}        
\usepackage{type1cm}        
\usepackage{makeidx}         
\usepackage{graphicx}        
\usepackage{multicol}        
\usepackage[bottom]{footmisc}

\makeindex             

\usepackage{paralist, amssymb}

\usepackage[latin1]{inputenc}
\newcommand{}{é}
\newcommand{}{\'E}
\newcommand{}{\"{\i}}
\newcommand{}{ô}
\newcommand{}{\"o}
\newcommand{}{\`u}
\newcommand{}{ü}
\newcommand{ç}{\'A}
\newcommand{}{\'a}
\newcommand{}{\'i}
\newcommand{}{\'u}
\newcommand{}{\'o}

\begin{document}

\title*{General Relativistic Simulations of the Collapsar Scenario}
\author{Nicolas de Brye, Pablo Cerd-Durn, Miguel çngel Aloy and Jos Antonio Font}
\authorrunning{N. de Brye et al.}
\institute{Nicolas de Brye, Pablo Cerd-Durn, Miguel çngel Aloy and Jos Antonio Font \at Departamento de Astronoma y Astrofsica, Universidad de Valencia, 46100 Burjassot (Valencia), Spain, \email{nicolas.de-brye@uv.es}}
%
%
\maketitle

\abstract{We are exploring the viability of the collapsar model for long-soft gamma-ray bursts. For this we perform state-of-the-art general relativistic hydrodynamic simulations in a dynamically evolving space-time with the CoCoNuT code. We start from massive low metallicity stellar models evolved up to core gravitational instability, and then follow the subsequent evolution until the system collapses forming a compact remnant. A preliminary study of the collapse outcome is performed by varying the typical parameters of the scenario, such as the initial stellar mass, metallicity, and rotational profile of the stellar progenitor. 1D models (without rotation) have been used to test our newly developed neutrino leakage scheme. This is a fundamental piece of our approach as it allows the central remnant (in all cases considered, a metastable high-mass neutron star) to cool down, eventually collapsing to a black hole (BH). In two dimensions, we show that sufficiently fast rotating cores lead to the formation of Kerr BHs, due to the fall-back of matter surrounding the compact remnant, which has not been successfully unbounded by a precedent supernova shock.}

\section{Introduction}
\label{sec:1}

Gamma-ray bursts (GRBs), routinely recorded by means of onboard satellite observatories, are one of the most luminous astrophysical events known. As they do not repeat, they must be catastrophic events. The tremendous energy and high variability at stake hint at the long GRBs to be sequels of the formation process of hyper-accreting stellar mass BHs. Thus we will focus on modelling massive rotating progenitor stars collapsing to BH and developing a thick accretion disk in their vicinity.

The art of creating a collapsar model is based on selecting the physics playing an allegedly key part in the process. Still, it has to be simplified with approximations to be able to simulate it on reasonable CPU times. The final fate of a massive star is a quite complex process, whose prevailing conditions involve the fundamental interactions of nature. These are 
\begin{inparaenum}[(i)]
\item gravity, modelled with general relativity (GR), approximated by the conformally flat condition \cite{Is78,Wal96,Cal11}, which is exact for spherical symmetry;
\item the weak interaction between baryonic matter and leptons, modelled with selected deleptonization processes, that are approximated with a parametric fit \cite{Li05} for the collapse phase, and an energy-gray leakage scheme for the post-bounce evolution; 
\item the strong nuclear interaction between baryonic particles, for which we employ a microphysical equation of state (EoS) \cite{LS91}; and 
\item electromagnetism, not included in this work, that would be modelled with the MHD theory in the GR framework, and whose implications are promising for explaining the stellar matter accretion energy transformation into the GRB jet kinetic energy.
\end{inparaenum}

Technically, as the central singularity {\it begins} to form, one needs to prescribe a procedure to follow the space-time hypervolume which will end up inside of the event horizon. Thus, we need to implement an apparent horizon (AH) finder.

Finally, we perform a number of 2D simulations with CoCoNuT \cite{Dal05}, in order to include rotation (breaking the initial spherical symmetry). We will show that rotating models naturally develop convective motions as well as a handful of hydrodynamic instabilities, such as the standing accretion shock instability (SASI). Hereafter, we briefly describe the deleptonization schemes employed and discuss some preliminary spherical symmetry and 2D equatorial symmetry results. 

\section{Deleptonization Schemes}
\label{sec:2}

The deleptonization schemes employed make the neutrino physics enter the local hydrodynamics conservation equations in the form of source terms: the nuclear composition change rate, and the energy-momentum exchange between the fluid and the radiative neutrino field.

The pre-supernova (SN) initial model starts collapsing due to its baryon self-gravity. In this hot dense matter, the weak interaction processes timescale becomes smaller than the dynamical timescale, and the core begins deleptonizing mainly by electron captures, which yields a copious amount of neutrinos that escape out of the core. As the collapse proceeds and the density rises ($\sim\rm 4\,10^{11}\,g\,cm^{-3}$), these neutrinos become trapped, forming a neutrinosphere a few milliseconds before core bounce. In the trapped core region, neutrinos thermalize by scattering, and diffuse out, a process that we include with the Liebendrfer prescription \cite{Li05} that reproduces the consequences of the delicate neutrino thermalization-diffusion process. A fit of the electron fraction as a function of the density (obtained in spherical symmetry simulations including full neutrino transport) permits deleptonizing in a reasonably realistic way up to bounce. The electron fraction loss and entropy changes are deduced from this fit.

Once the saturation density ($\rm 2\,10^{14}\,g\,cm^{-3}$) of nuclear matter is reached, the strong nuclear interaction suddenly turns matter more incompressible, and a shock wave forms. In this post-bounce phase, the previous Liebendrfer fit cannot reproduce the deleptonization, and a neutrino leakage scheme based on \cite{OO10,RL03,Ru96} serves as another neutrino cooling approximation. It relies upon splitting up the stellar core interior in two regions: one denser, where the neutrino diffusion timescale is longer than the dynamical timescale (neutrinos are trapped and reach $\ubeta$-equilibrium at center, i.e. $\dot Y_e=\partial_t Y_e=0$), and another less dense beyond the neutrinosphere where neutrinos stream out freely. In the intermediate semi-transparent region, an empirical opacity-based interpolation allows us approximating the neutrino transport. Of course, this approximation to the true (much more costly) neutrino radiative transport shall be regarded as a first step towards implementing more elaborated schemes. 

The neutrino interactions treated in this leakage, exchanging energy and/or lepton number, are charged current $\ubeta$-processes on nucleons and nuclei, neutral current elastic scattering on nucleons and nuclei, and thermal neutrino-pair production-absorption with electron-positron pair and transversal plasmon decay. The neutral current neutrino-electron inelastic scattering cannot be properly included in this energy-gray leakage scheme. However, this process is only important before the shock breaks through the neutrinosphere, i.e. in a dynamical phase where we are using the Liebendrfer prescription, where the aforementioned microphysics is properly included. The opacity is mainly due to scattering in this first phase, and then due to absorption-emission in the second phase.

\section{Results and Discussion}
\label{sec:3}

\begin{figure}
\vskip -10pt
\includegraphics[scale=.5, trim=20 0 0 0, clip=true]{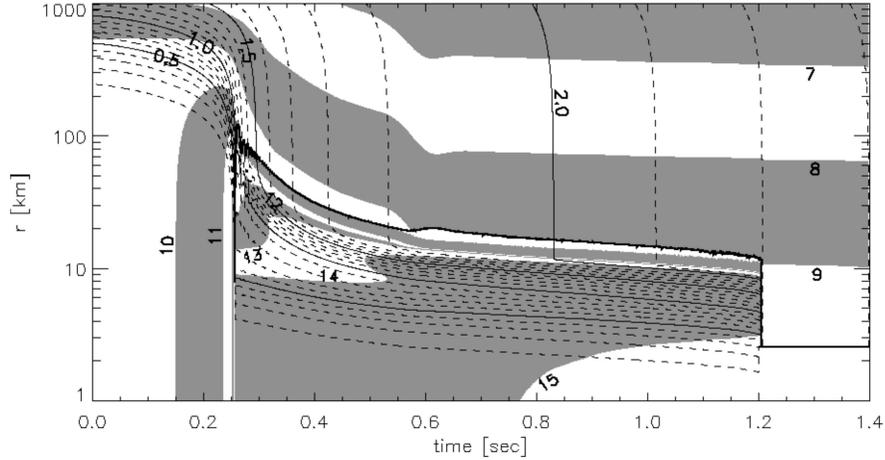}
\vskip -10pt
\caption{Time evolution, for model s40 on the first 1000 km, of the shock radius ({\it thick solid line}), the isopycnal positions of $10^7$--$10^{15}\,$g\,cm$^{-3}$ ({\it grey strip contours}), and the equal enclosed mass positions of 0.1--$\rm 2.0\,M_\odot$ in steps of $\rm 0.1\,M_\odot$, and 2.0--$\rm 2.15\,M_\odot$ in steps of $\rm 0.05\,M_\odot$ ({\it solid and dashed lines}). The bottom-right corner corresponds to the excised region within the AH}
\label{fig:1}
\vskip -10pt
\end{figure}

\begin{minipage}{\linewidth}
We have improved the leakage scheme of \cite{OO10} to match GR simulations of the G15 model \cite{Lal05} with full Boltzmann transport up to 250 ms after bounce. We have performed simulations of several progenitor models of \cite{Wal02} to test for mass and metallicity effects. Models with initial iron core mass just above the Chandrasekhar mass form an AH very late (between 3 to 5 s after core bounce). The reason is the very small accretion rate onto the newly formed proto-neutron star (PNS). Nevertheless, over such long periods, the validity of our neutrino transport approximation is doubtful, since a proper transport scheme may well yield a successful SN explosion. Thus our models predictions regarding late BH formation shall be taken with special care. We have checked that, in agreement with previous studies \cite{OO11}, the heavier the core, the faster the AH forms. In the sample of initial models at hand, the heaviest cores correspond to stars with the lower metallicity. Indeed, the observed trend confirms that the most likely progenitor stars producing collapsars are the low metallicity ones, which correlates with those having the most massive iron cores. It is also worth mentioning that our simulations did not lead to direct collapse to a BH even for the most massive models of $\rm 75\,M_{\odot}$; all BH formation happened by post-bounce accretion, and driven by neutrino-cooling. Figure~\ref{fig:1} shows the 1D space-time evolution of the pre-SN initial model s40, employing the LS180 EoS during the collapse phase up to 0.257 s, the PNS phase to 1.206 s, and the BH phase.
\end{minipage}

After having optimized our leakage scheme, we are currently obtaining a grid of numerical models with an ad hoc rotational profile on top of 1D stellar progenitors. Preliminary results show that convection and SASI develop in the stellar cores, delaying the AH formation. However, a Kerr BH eventually forms, and the centrifugal barrier halts the accretion onto it, yielding the formation of a thick accretion disk. More detailed results of this process will be subject of a future publication.

\begin{acknowledgement}
This work was supported by the Spanish Ministry of Education (AYA2010-21097-C03-01), the European Research Council (STG-ERC-259276-CAMAP) and the Generalitat Valenciana (PROMETEO-2009-103).
\end{acknowledgement}

%

\end{document}